\documentstyle[prc,preprint,aps]{revtex}
\def\sd #1 #2 #3 #4 {\left| \begin{array}{ccc} #1 & #2 \\
#3 & #4 \end{array}\right| }
\def\tj #1 #2 #3 #4 #5 #6 {\left( \begin{array}{ccc} #1 & #2 & #3 \\
#4 & #5 & #6 \end{array}\right) }
\def\sj #1 #2 #3 #4 #5 #6 {\left\{ \begin{array}{ccc} #1 & #2 & #3 \\
#4 & #5 & #6 \end{array}\right\} }
\def\nj #1 #2 #3 #4 #5 #6 #7 #8 #9
{\left\{ \begin{array}{ccc} #1 & #2 & #3 \\
#4 & #5 & #6 \\ #7 & #8 & #9 \end{array}\right\} }
\def\ls #1 #2 #3 #4 #5 #6 #7 #8 #9
{\left[ \begin{array}{ccc} #1 & #2 & #3 \\
#4 & #5 & #6 \\ #7 & #8 & #9 \end{array}\right] }
\makeatletter

\long\def\@caption#1[#2]#3{
  \begingroup
    \@parboxrestore
    \small
    \baselineskip=12pt
    \advance\leftskip by 1cm
    \advance\rightskip by 1cm
    \@makecaption{\csname fnum@#1\endcsname}{\ignorespaces
     {#3}}\par
  \endgroup}
\makeatother
\begin{document}
\title{Coulomb Displacement Energy and the Low-Energy Astrophysical
$  S_{17}  $  Factor for the $^{7}$Be(p,$\gamma$)$^{8}$B Reaction}
\author{B. A. Brown}
\address{National Superconducting Cyclotron Laboratory and
Department of Physics and Astronomy\\
Michigan State University, East Lansing, MI 48824-1321}
\author{A. Cs\'ot\'o}
\address{National Superconducting Cyclotron Laboratory\\
Michigan State University, East Lansing, MI 48824-1321}
\author{R. Sherr}
\address{Department of Physics, Princeton University, Princeton,
New Jersey 08544}
\maketitle
\begin{abstract}
The relationship between the Coulomb displacement energy for the A=8,
J=2$^{ + }$, T=1 state and the low-energy astrophysical $  S_{17}  $
factor for the $^{7}$Be(p,$\gamma$)$^{8}$B
reaction is discussed. The displacement energy is interpreted in a
particle-hole model.
The dependence of the particle
displacement energy on the potential well geometry is investigated and
is used to relate the particle displacement energy to the
rms radius
and the asymptotic normalization of the valence proton
wave function in $^{8}$B. The asymptotic normalization is used to
calculate
the astrophysical $  S_{17}  $ factor for the $^{7}$Be(p,$\gamma$)
reaction.
The relationship to the
$^{7}$Li(n,$\gamma$) reaction, the $^{8}$B quadrupole moment,
radial density, and break-up
momentum distribution are also
discussed.
\end{abstract}

\section{Introduction}

Measurements of high energy solar neutrinos in the Homestake\,$^{1}$ and
Kamiokande II and III\,$^{2}$ experiments have found a significantly
smaller number of solar neutrinos compared to that expected from the
standard solar model.\,$^{3}$ The $^{8}$B $\beta^{ + }$ decay is the
main
source of these these high energy
solar neutrinos. $^{8}$B is formed by the
$^{7}$Be(p,$\gamma$)$^{8}$B reaction
at a center of
mass energy of about 20 keV. Its cross section is conventionally
expressed in terms of the
$  S_{17}  $ factor.\,$^{3}$ $  S_{17}  $ is known only from an
extrapolation of data at energies above 100 keV. Also the highest
precision data disagree, with those of Parker\,$^{4}$
and Kavanagh et al.\,$^{5}$
being about 30 percent higher than those of Filippone et al.\,$^{6}$ and
Vaughn et al..\,$^{7}$ This has lead to the investigation of other ways
to
determine the $  S_{17}  $ factor such as the Coulomb
dissociation,\,$^{8}$ as
well as other properties of $^{8}$B which are indirectly related to the
$  S_{17}  $ factor such as its total reaction cross section,\,$^{9}$
quadrupole moment,\,$^{10}$ and break-up momentum distribution.\,$^{11}$

Much of the theoretical work has been based upon using a potential model
to generate the single-particle wave function for the most loosely bound
proton, and then combining this with a shell-model calculation of the
spectroscopic factor to obtain the asymptotic normalization of the wave
function. This current work was initiated by the observation that the
potential
model parameters (e.g. the radius $  R  $ and diffuseness $  a  $ of the
Woods-Saxon potential) used in previous calculations
are taken from some ``standard" sets based
upon nucleon-nucleus scattering optical potential analyses, and that
these ``standard" parameters were not obviously appropriate to the
particular case of $^{7}$Be plus protons. We thus investigated the
extent to which the potential
parameters could be determined from
the displacement energy of the A=8, J$^{ \pi }$=2$^{ + }$, T=1
state as given by the binding energy difference between the $^{8}$B and
$^{8}$Li
ground states. In addition, we also
examined several quantities of related interest including the
$^{7}$Li(n,$\gamma$) cross section, the $  Q  $ moment,
the density distributions, and the momentum distributions.

\section{Coulomb displacement energy}

The relationship between the
root-mean-square (rms) radius of the valence proton and the displacement
energy is well known.\,$^{12,13}$ Qualitatively, the larger the rms
radius the
smaller the displacement energy since the valence proton is further
from the core protons. This leads to the Thomas-Ehrman effect in which
loosely bound valence nucleons have a relatively
smaller displacement energy compared to more tightly bound nucleons.
In addition, for a fixed binding energy, nucleons in a low $\ell$ state
have a smaller displacement energy compared to those in a high $\ell$
state
because the centrifugal barrier for high $\ell$ results in a smaller
rms radius. For heavier nuclei (A$>$16)
with a relatively simple
shell-model configuration, many quantitative calculations of the
displacement energy have been carried out. At the beginning of the
0d1s shell (A=17, T=1/2) and the 0f1p shell (A=41, T=1/2), the observed
displacement energy
is found to be about 10 percent larger than that
obtained from calculations which involve only the lowest shell-model
configuration together with the Coulomb interaction. These cases
include those for high $\ell$ values whose Coulomb shift is not
very sensitive to the potential-well geometry.
This
``Nolen-Schiffer" (NS) anomaly appears not to have a simple explanation
but is due to a combination of core-polarization, high-order
configuration
mixing, and charge-asymmetric strong interactions effects.
The systematics
of the displacement energies of light nuclei can also be
semi-quantitatively
accounted for with a simple potential-well geometry.\,$^{14}$

In this paper we apply what has been learned about the displacement
energy systematics to constrain the shape of the potential for
the $^{7}$Be plus valence proton system. The A=8, T=1 displacement
energies
are more complicated than the T=1/2 systems usually considered,
since at least two nucleons are involved. However,
the 0p shell-model wave functions predict a relatively simple structure
for A=8, T=1. The spectroscopic
factors (C$^{2}$S) for (A=8, 2$^{ + }$, T=1) $\rightarrow$
(A=7, 3/2$^{-}$, T=1/2) and
(A=9, 3/2$^{-}$, T=1/2) $\rightarrow$ (A=8, 2$^{ + }$, T=1)
are experimentally and theoretically both near
unity.\,$^{15,16}$
Thus, $^{8}$B can be considered as a
proton-particle neutron-hole configuration relative to $^{8}$Be. The
mirror nucleus
$^{8}$Li is a proton-hole neutron-particle configuration. $^{8}$Be is
itself
``deformed" and thus exits in both the ground state and
2$^{ + }$ first excited state relative to the particle-hole
configuration.

The
displacement energy of the particle-hole state is
simply the sum of the particle and hole displacement energies
(the Coulomb particle-hole interaction is zero):

$$
\Delta _{8} = \Delta _{h} + \Delta _{p},       \eqno({1})
$$

\noindent
  where $  \Delta _{8}  $ is the A=8, T=1 displacement energy (the
binding energy difference between $^{8}$Li and $^{8}$B). In the
simplest approximation $  \Delta _{h}  $ is the
A=7, 3/2$^{-}$, T=1/2 displacement energy and $  \Delta _{p}  $ is the
A=9, 3/2$^{-}$, T=1/2 displacement
energy. The experimental values are $  \Delta _{7}  $=1.645 MeV and
$  \Delta _{9}  $=1.851 MeV which
gives $  \Delta _{7}  +  \Delta _{9}  $=3.496 MeV compared to the
experimental A=8 value
of $  \Delta _{8}  $=3.540 MeV. So the simplest model works rather well.
There are however several reasons why this is not exact. One is
that the actual shell-model configuration is a little more complicated
than just particle-hole. Also the Thomas-Ehrman effects may
be significant since the proton is unbound by
by 186 keV in $^{9}$B and bound by 138 keV in $^{8}$B.

Since we are particularly interested in the properties of the
valence proton in $^{8}$B we will proceed as follows. First Eq.\ (1)
will be modified to read

$$
\Delta _{8} = \Delta _{h} + \Delta '_{p} + \Delta _{sm},
\eqno({2})
$$

\noindent
  where $  \Delta _{sm}  $ will take into account the 0p shell-model
structure beyond particle-hole.
Since the hole state is relatively tightly bound, the Thomas-Ehrman
shift will be small and we will take
$  \Delta _{h}  $ = $  \Delta _{7}  $ = 1.645 MeV.
$  \Delta '_{p}  $ indicates the displacement energy for the
particle
in $^{8}$B which may differ from $  \Delta _{p}=\Delta _{9}  $ because
of the Thomas-Ehrman
shift.

The shell-model correction $  \Delta _{sm}  $ was calculated by using
the
Coulomb plus charge asymmetric interaction of Ormand and Brown\,$^{17}$
within a full 0p shell-model basis.
The matrix elements were calculated with harmonic-oscillator radial
wave functions.
The results for the
displacement energies are: $  \Delta _{h}=\Delta _{7}=1.719  $ MeV, $
\Delta _{8}=3.656  $
MeV and $  \Delta _{p}=\Delta '_{p}=\Delta _{9}=1.873  $ MeV. This
gives $  \Delta _{sm}=64  $ keV.
Since harmonic-oscillator radial wave functions are used the shell-model
calculation should not necessarily be
in good agreement with experiment, however,
it should provide an estimate for the  $  \Delta _{sm}  $ correction.
Thus the displacement energy of the A=8 particle state is

$$
\Delta '_{p} = \Delta _{8} - \Delta _{h} - \Delta _{sm} = 1.831 \, {\rm
MeV} .       \eqno({3})
$$

We are interested on how $  \Delta '_{p}  $ depends upon the potential
geometry,
and on how the potential geometry effects the rms radius
and the astrophysical $  S_{17}  $ factor. To investigate this we
will calculate the direct part of the Coulomb shift in a Woods-Saxon
geometry. This means that we calculate the single-particle binding
energy with and without the one-body Coulomb potential and take
the difference to obtain the Coulomb displacement
energy.
The Woods-Saxons potential has the usual form of
central plus spin-orbit plus Coulomb terms. The central potential
has the form

$$
V(r) = V_{o}\{1+{\rm exp}[r-R]/a]\}^{-1},       \eqno({4})
$$

\noindent
  where $  R  $ is the radius, $  a  $ is the diffuseness.
We use the reduced mass in the kinetic energy operator.
The spin-orbit potential is the usual derivative form with the same
geometry as the central (we will show below that the spin-orbit
potential is not important for our analysis).
The Coulomb potential is obtained from the
density distribution of four tightly-bound protons (two in 0s and two in
0p) obtained with another Woods-Saxon potential which is constrained to
reproduce the experimental rms charge radius\,$^{18}$
 of 2.52 fm for $^{9}$Be.

We thus investigate the dependence of the Coulomb energy, the rms
radius and the astrophysical $  S_{17}  $ factor on the radius $  R  $
and
the diffuseness $  a  $ of the central potential.
In all cases we vary $  R  $ and $  a  $ and fix $  V_{o}  $
in order to reproduce the proton separation energy of 138 keV.
We calculate the
direct Coulomb displacement energy $  \Delta _{o}  $ and relate it to
$  \Delta '_{p}  $ by making several additional corrections, all of
which
are relatively small:

$$
\Delta '_{p} = \Delta _{o} + \Delta _{ex} + \Delta _{so} + \Delta _{vp}
+ \Delta _{np}
+ \Delta _{NS}.       \eqno({5})
$$

\vspace{ 12pt}
\noindent
  $  \Delta _{ex}  $ is the exchange correction. A value of
$  \Delta _{ex}  $=--125 keV
was obtained
from the harmonic-oscillator
shell-model calculation for $  \Delta _{9}  $ by comparing the results
with and without the Coulomb exchange
terms. The next three terms are the
relativistic spin-orbit correction (Eq.\ 21 in Ref 12), the
vacuum polarization correction (Eq.\ 4.23 in Ref 13) and the
proton-neutron mass difference correction (Eq.\ 4.29 in Ref 13),
respectively.
Our estimates for these are $-$30 kev, 12 keV and
14 keV, respectively. The Nolen-Schiffer correction, $  \Delta _{NS}  $
is perhaps the most uncertain. For A=17 and A=41, $  \Delta
_{NS}/\Delta _{o}  $ is
about 0.10 and we see no reason why it should differ much from this
for the beginning of the 0p shell. Since $  \Delta _{o}  $ is about 1.8
MeV
we will take $  \Delta _{NS}  $=180 keV. Thus we arrive at the
empirical value of

$$
\Delta _{o}({\rm empirical}) = 1.780\, {\rm MeV},       \eqno({6})
$$

\noindent
  with which we will compare our calculations for the direct Coulomb
shift of the valence particle.

We show in Table I the results for $\Delta_{o}$, the rms proton radius
and the
density of the valence proton at $  r=10  $ fm, $  \rho (10\, {\rm fm})
 $, as a function of
$  R  $ and $  a  $.
Beyond the influence of the strong interaction (about 6 fm) the shape of
the radial wave function is entirely and uniquely determined by the
Coulomb plus centrifugal potentials. The only quantity which depends on
the potential is the asymptotic normalization as represented, for
example, by the value of $\rho$ at $  r=10  $ fm. Our $\rho$(r) is
defined by the
normalization:

$$
4\pi \displaystyle\int \rho (r) r^{2} \,dr = 1.       \eqno({7})
$$

\noindent
  In order to show the
correlation between $\Delta_{o}$, rms and $  \rho (10\, {\rm fm})  $ we
plot values obtained for
these quantities in pairs in Figs.\ 1 ($\Delta_{o}$ vs rms), 2
[$\Delta_{o}$ vs
$  \rho (10\, {\rm fm})  $]
and 3 [rms vs $  \rho (10\, {\rm fm})  $]. We find that there is strong
correlation between them, which implies that from a knowledge of any one
of them (in particular $\Delta_{o}$) we can infer a range of values for
the
other two.

The results in Table I were obtained for a 0p$_{3/2}$ valence particle.
For $  R  $=2.4 fm we show the results with and without the
spin-orbit potential, and from this comparison it is clear that
spin-orbit potential is not important (as long as $  V_{o}  $ is fixed
from
the 138 keV separation energy). Thus to a good approximation, our
results apply to both 0p$_{3/2}$ and 0p$_{1/2}$.

The results shown in Figs.\ 1$-$3 show that there is a rather narrow
band of points which relate the quantities of interest in the
framework of the Woods-Saxon potential model. From Fig.\ 1 one
finds that the value of $\Delta_{o}$=1.780 MeV corresponds to a narrow
range of rms valence radii from 4.5 to 4.7 fm, and from
Fig.\ 2 one finds that the same $\Delta_{o}$ corresponds to
$  \rho (10\, {\rm fm})  $ values in the range 7.2 to 8.0 fm$^{-3}$.
The correlation between the rms radius and $\rho$ shown in Fig.\ 3
is even more parameter independent.

The potential parameter values required fall into three sets with
$  (R, a)  $ values of about (2.8, 0.51 fm) (set A), (2.4, 0.69 fm)
(set B),
and (2.0, 0.81 fm) (set C). Or analysis is consistent with
any of these sets or any interpolation between them. These are
compared in Table II to the parameters used in a number of
other potential model calculations for $  S_{17}  $.
The results for $\Delta_{o}$, rms and $  \rho (10\, {\rm fm})  $ for
the potentials
in Table II are shown by the labeled crosses in Figs.\ 1-3.
The results with
Tombrello parameters (T) are close to ours. They were obtained from the
optical model analysis of 180 MeV proton scattering on Li and Be by
Johansson et al..\,$^{19}$ The Johansson parameters were subsequently
used by Aurdal\,$^{20}$ and Robertson.\,$^{21}$ Parameter set I from
Barker is also close to ours. But his modifications of
the parameters needed to reproduce the $^{7}$Li(n,$\gamma$) cross
section (see the
discussion below) are clearly outside of our range.

\section{Results for the astrophysical $  S_{17}  $ factor}

The astrophysical $  S_{17}  $ factor is related to $  \rho (10\, {\rm
fm})  $ in the
following way.
At very low energies ($<$ 50 keV) the $^{7}$Be(p,$\gamma$)$^{8}$B
capture
cross section is dominated by the E1 transition between the $^{7}$Be$+
p  $
scattering states and the $^{8}$B ground state,\,$^{22}$ and it is
almost
exclusively determined by contributions coming form the external part
of the scattering and bound state wave functions. The asymptotic
behavior
of the scattering states is uniquely defined (the phase shifts being the
hard sphere phase shifts -- practically zero), and the asymptotic
part of the wave function, which describes the $^{7}$Be$+  p  $ relative
motion in the bound state $^{8}$B, is proportional to the fixed
Whittaker function,

$$
   \psi ^{I}_{^{8}{\rm B}}(r)=\bar c_{I} \frac{W^{+}_{\eta,\ell
}(kr)}{r},
    \hskip 0.5cm r \rightarrow\infty,       \eqno({8})
$$

\noindent
  where $  I=1,2  $ is the channel spin, $  r  $ is the radial distance
between
$^{7}$Be and $  p  $, and $  k  $ is the wave number corresponding to
the $^{8}$B
binding energy relative to the $^{7}$Be$+  p  $ threshold.
$  \bar c_{I}  $ are the constants which are required to normalize
the Whittaker function to the asymptotic $^{8}$B wave function.
Thus, at low
energies the astrophysical $  S_{17}  $ factor of the capture reaction,

$$
S_{17}(E)=\sigma(E)E\exp\left [2\pi\eta (E)\right ],       \eqno({9})
$$

\noindent
  depends only on $  \bar c  $.\,$^{23,24}$
(here $  \eta =e^{2}Z_{1}Z_{7}/\hbar \nu   $ with $  Z_{1}=1  $,
and $  Z_{7}=4  $ is the Sommerfeld parameter).
{}From the hard sphere scattering states one obtains:

$$
S_{{}17}(20{\rm keV})=36.5(\bar c_{1}^{2}+\bar c_{2}^{2}).
\eqno({10})
$$

\noindent
  (The uncertainty coming mainly from the fact
that the nucleon mass is not well defined in nonrelativistic
quantum mechanics, is 1$-$2 percent. The value of the $  S  $ factor at
zero energy is roughly 0.4 eV-barn higher than at 20 keV.)
Using this formula we can express $  S_{{}17}  $
(in units of eV-barns)
in terms of the valence
proton density at any given asymptotic radius, e.g.\ at 10 fm,

$$
S_{17}(20 \, {\rm kev}) = 2.99^{.}10^{6}\,\rho _{3/2}(10\, {\rm
fm})\,\,S_{3/2}\,\,
[(\alpha _{1,3/2}+\gamma \,\alpha _{1,1/2})^{2} + (\alpha
_{2,3/2}+\gamma \,\alpha _{2,1/2})^{2}],       \eqno({11})
$$

\noindent
  were the $\alpha$ coefficients are determined from the transformation
between the $^{7}$Be($  J_{i}  $) $+$ $\ell_{j}$ = $^{8}$B($  J_{f}  $)
coupling and the
channel-spin coupling:

$$
[J_{i}\otimes \ell _{j}]^{J_{f}} =
\displaystyle\sum _{I} \alpha _{I,j} [(J_{i}\otimes 1/2)^{I}\otimes
\ell ]^{J_{f}},       \eqno({12})
$$

\noindent
  and $\gamma$ is given by the ratio:

$$
\gamma  = \frac{[\theta _{1/2}\,\,\psi _{1/2}(10\, {\rm fm})]}{[\theta
_{3/2}\,\,
\psi _{3/2}(10\, {\rm fm})]}.       \eqno({13})
$$

\noindent
  In Eq.\ (13)  $  \theta _{j}  $ is the $  n=0  $, $\ell$=1
spectroscopic amplitude and in Eqs.\ (11) $  S_{j}  $ is the
spectroscopic
factor. The amplitudes $  \theta _{j}  $ are given by the
reduced matrix elements\,$^{25}$ of the creation operator, $  a^{+}  $:

$$
\theta _{j}(J_{i},J_{f}) = \frac{\langle {\rm ^{8}B} ,J_{i}\mid \mid
a^{+}_{j, {\rm proton}}\mid \mid {\rm ^{7}Be} ,J_{f}\rangle }
{\sqrt{2J_{i}+1}\, },       \eqno({14})
$$

\noindent
  When $\theta_{j}$ is given without its $  J_{i},J_{f}  $ arguments as
in
Eq.\ (13), it corresponds to the $^{8}$B ground state ($  J_{i}=2  $)
to $^{7}$Be ground state ($  J_{f}=3/2  $) value.
The spectroscopic factors take into account the additional
center of mass correction factor,\,$^{26}$ $  [A_{i}/(A_{i}-1)]=8/7  $:

$$
S_{j} = \frac{A_{i}}{A_{i}-1} \theta ^{2}_{j},       \eqno({15})
$$

\noindent
  In Eq.\ (13), $  \psi (10\, {\rm fm})  $ is the radial amplitude at $
 r=10  $  fm:

$$
\rho _{j}(10\, {\rm fm}) = \psi _{j}^{2}(10\, {\rm fm}).
\eqno({16})
$$

\noindent
  In our case where $\ell$=1, $  J_{i}=3/2  $, $  J_{f}=2  $ and $
I=1,2  $,
the transformation coefficients are
$\alpha_{1,3/2}$=$\alpha_{2,3/2}$=$\alpha_{2,1/2}$=1/$\sqrt{2}\,$ and
$\alpha_{1,1/2}$=--1/$\sqrt{2}\,$. In addition, to a good approximation,
$  \psi _{1/2}(10\, {\rm fm})=\psi _{3/2}(10\, {\rm fm})=\psi (10\,
{\rm fm})  $.
Hence Eq.\ (11) simplifies to:

$$
S_{17}(20\, {\rm kev}) = 2.99^{.}10^{6}\,\rho (10\, {\rm fm})\,S,
\eqno({17})
$$

\noindent
  where $  S=S_{3/2}+S_{1/2}.  $

In Table III we give the values of $  \theta _{j}  $ obtained from
0p shell model calculations with a variety of interactions
which are appropriate for the lower part of the 0p shell. These
amplitudes are quite stable with respect to a reasonable range
of interactions.

In Table IV we compare the theoretical
spectroscopic factors with those obtained from reaction data\,$^{15}$
for states of $^{8}$Li as well as those extracted from the
observed widths of unbound states in $^{8}$Li and $^{8}$B.\,$^{27}$ The
spectroscopic factors for the unbound states are obtained
from $  \Gamma _{exp} = S \, \Gamma _{sp}  $ where $\Gamma_{sp}$ is the
single-particle width for a resonance at the experimental
separation energy in the potential geometry B ($  R=2.4  $ fm and
$  a=0.69  $ fm). The decay data are observed to be in excellent
agreement with theory. The reaction spectroscopic factor for the 2$^{ +
}$ state
is low with respect to theory, however, it depends upon the potential
parameters used for the DWBA calculations (we have not attempted to
repeat the DWBA calculations). The reaction spectroscopic
factor for the 1$^{ + }$ state is also low with respect to the decay
spectroscopic factor of the mirror state.
The data in Table IV support
the 0p shell-model calculation for $  S  $ and the present potential
parameters.

Combining with our results of $  S=1.15\pm 0.05  $ and
$  \rho (10\, {\rm fm})=(7.7\pm 0.4)^{.}10^{-6}  $ fm$^{-3}$ we obtain
$  S_{17}(20\, {\rm keV})=26.5\pm 2.0  $ eV-barns.

\section{Relationship to the $^{7}$Li(n,$\gamma$)$^{8}$Li cross section}

Barker\,$^{28}$ has pointed out that standard potential models tend to
overestimate the experimentally well-determined low-energy
$^{7}$Li(n,$\gamma$)$^{8}$Li cross section. He argued, that
one should modify either the potential parameters or the spectroscopic
factor to get agreement with experiment. To study this issue, we
performed calculations for the $^{7}$Li(n,$\gamma$)$^{8}$Li reaction.

Because there is no Coulomb barrier in this reaction, the inner parts of
the wave functions have the same importance as the asymptotic parts.
Thus
the cross section does not depend solely on the asymptotic normalization
of the bound state wave function. For the $^{8}$Li bound state we used
the
same potential parameters as for $^{8}$B, except a  change in the
potential depth to get the exact neutron separation energy of 2.033 MeV.
For the $^{7}$Li$+  n  $ scattering states, we modify the potentials to
reproduce
the experimental scattering lengths of the $  I=1  $, and $  I=2  $
channel
spin states, respectively.\,$^{28}$

The thermal $^{7}$Li(n,$\gamma$)$^{8}$Li cross section in our model,
using
potentials A, B, and C, are 78, 80 and 83 mb, respectively. The
experimental thermal cross section is 45.4$\pm$3.0 mb.\,$^{27}$
Thus we also obtain an overestimation of the
$^{7}$Li(n,$\gamma$)$^{8}$Li cross section as did Barker. He concluded
that
either the potential parameters or the spectroscopic factor has to be
changed in order to agree with the experiment, and that
these changes would
bring the $^{7}$Be(p,$\gamma$)$^{8}$B $  S_{17}(20\, {\rm keV})  $
factor down to 16-17 eV-barns.
The modifications to the potential are very large (R is changed
from 2.39 to 0.53 fm or $  a  $ is change from 0.65 fm to 0.27 fm) and
we can see from Figs.\ 1$-$2 that these large changes are inconsistent
with the Coulomb displacement energy. Thus we
can exclude the
possibility of radically changing the potential parameters. In the
spirit of Ref 28, the only remaining possibility would be the
reduction of the spectroscopic factor to about 0.71. But given
the general agreement we obtain for the decay widths in Table IV,
such a large change in the spectroscopic factor seems unreasonable.
In fact such a drastic change
would question the adequacy of the potential model itself.

We would like to point out that the discrepancy in the
$^{7}$Li(n,$\gamma$)$^{8}$Li cross section could be resolved in a way
which
does not affect the $^{7}$Be(p,$\gamma$)$^{8}$B cross section. As
mentioned,
contrary to $^{7}$Be(p,$\gamma$)$^{8}$B, the inner part of the wave
functions are important in the case of $^{7}$Li(n,$\gamma$)$^{8}$Li.
Although the reproduction of the scattering lengths fixes the external
part of the scattering wave functions, the internal, off-shell, part is
not well-constrained. For instance, if the inner node of the wave
function were somewhat further outside than in the potential model, this
would bring the $^{7}$Li(n,$\gamma$)$^{8}$Li cross section down. To
illustrate that the node position is not well-defined, we show in Fig.\
4
the inner part of the
$  I=2  $ scattering wave function of the standard potential (BI) of
Barker\,$^{28}$ (solid line) together with the scattering state
obtained
from the cluster model of Ref 29 (dashed line) at
$  E_{CM}  $=10 keV. This change in the off-shell behavior is enough
to reduce the $^{7}$Li(n,$\gamma$)$^{8}$Li cross section considerably.
In
fact, the dashed line of Fig.\ 4, together with the bound state of the
standard potential of Barker\,$^{28}$
results (after a $  1/v  $ extrapolation from 10 keV)
in a thermal cross section of 46.3
mb, which is close to the
experimental value. We emphasize again, that this
modification in the off-shell behavior of the scattering wave functions
has no effect on the $^{7}$Be(p,$\gamma$)$^{8}$B cross section.

\section{Relation to the $^{8}$B quadrupole moment}

The quadrupole moment, $  Q  $, of $^{8}$B is related to the above
calculations
in the following way. $  Q  $ is proportional to the matrix element of
the
$  E2=r^{2}Y^{(2)}  $ one-body operator whose reduce matrix element is
given
by a summation over products of many-body matrix elements times
single-particle matrix elements\,$^{30}$:

$$
\langle J_{i}\mid \mid E2\mid \mid J_{i}\rangle  = \displaystyle\sum
_{j,j',t_{z}}
\frac{\langle J_{i}\mid \mid [a^{+}_{j,t_{z}}\otimes
\tilde{a}_{j',t_{z}}]^{(\lambda )}\mid \mid J_{i}\rangle
}{\sqrt{2\lambda +1}\, }
\langle j,t_{z}\mid \mid E2\mid \mid j',t_{z}\rangle ,       \eqno({18})
$$

\noindent
  where $\lambda$=2.
The $  t_{z}  $ indicates a sum over protons and neutrons.
By inserting a complete set of states\,$^{31}$ ($  J_{f}  $) of the $
A=7  $
system between
the $  a^{+}  $ and $  \tilde{a}  $, one can rewrite this as a sum over
all 0p
shell states of the A=7 system:

$$
\langle J_{i}\mid \mid E2\mid \mid J_{i}\rangle  = (2J_{i}+1)
\displaystyle\sum _{j,j',t_{z},J_{f}}
(-1)^{J_{f}+j'} \theta _{j,t_{z}}(J_{i},J_{f}) \theta
_{j',t_{z}}(J_{i},J_{f})
$$
$$
\times\,  \sj J_{i} J_{i} 2 j j' J_{f}
\langle j,t_{z}\mid \mid E2\mid \mid j',t_{z}\rangle ,       \eqno({19})
$$

\noindent
  where
the states $  J_{f}  $ are in $^{7}$Be for $  t_{z}  $=proton and are
in $^{7}$B for
$  t_{z}  $=neutron. The sum over $  J_{f}  $ can be broken down into
the
term coming from the ground state of $^{7}$Be (referred to as the
valence
term, $  vp  $) and all other terms coming from excited states in
$^{7}$Be and all states in $^{7}$B (referred to as 0p core proton and
neutron
terms, $  pcp  $ and $  pcn  $, respectively). The single-particle
matrix
elements are given by a geometrical term times the single-particle
mean-square radius,
$  \langle r^{2}\rangle   $.\,$^{30}$ The $  Q  $ moment can thus be
expressed in the form

$$
Q(^{8}{\rm B}) = -0.80\, \theta _{3/2} \theta _{1/2}
\langle r^{2}\rangle _{vp} e_{p}
+ 0.203\, \langle r^{2}\rangle _{pcp} e_{p} + 0.183 \,\langle
r^{2}\rangle _{pcn} e_{n}
$$
$$
= 0.187 \langle r^{2}\rangle _{vp} e_{p}
+ 0.203\, \langle r^{2}\rangle _{pcp} e_{p} + 0.183 \,\langle
r^{2}\rangle _{pcn} e_{n},       \eqno({20})
$$

\noindent
  where the CKI interaction was used for the spectroscopic factors.
The numerical coefficient $-$0.80 is purely geometrical.
The effective charges $  e_{p}  $ and $  e_{n}  $ take into account
the non-0p parts of the wave functions which include 0s to 0p and 0p to
0d1s proton excitations. For the remaining discussion we will
use values of
$  e_{p}=1.35e  $ and $  e_{n}=0.35e  $,\,$^{30}$ although we realize
that these
are approximate values and that they may depend upon the binding
energy.\,$^{32}$ Our calculations for the radial matrix elements
(with potential B) give
$  \langle r^{2}\rangle _{vp}  $ = 21.6 fm$^{2}$,
$  \langle r^{2}\rangle _{pcp}  $ = 8.1 fm$^{2}$,
and $  \langle r^{2}\rangle _{pcn}  $ = 7.8 fm$^{2}$, and hence we
obtain
$  Q  $($^{8}$B)  = 8.2 $  e  $\,fm$^{2}$, in reasonable agreement with
the experimental value\,$^{10}$ of 6.83$\pm$0.21 $  e  $\,fm$^{2}$. The
$  Q  $ moment for
$^{8}$Li can be obtained by interchanging the labels for $  p  $ and $
n  $ in
Eq.\ (20), and using our value of
$  \langle r^{2}\rangle _{vn}  $ = 14.0 fm$^{2}$, to obtain $  Q
$($^{8}$Li) = 3.47 $  e  $\,fm$^{2}$ which
is close to the experimental value\,$^{10}$ of
3.27$\pm$0.06 $  e  $\,fm$^{2}$. The effective charges
we use are approximate and it would very difficult to estimate the
them more quantitatively.
Even though our calculationed values are close
to the experimental values, it is
clear that the relationship to the valence rms radius and asymptotic
normalization is quite complicated.
In addition, we note that the $  Q  $ moment depends upon the
interference term $  \theta _{3/2}\theta _{1/2}  $ whereas the tail
density is
determined by the combination $  A^{2}_{3/2}+A^{2}_{1/2}  $ which is
dominated by $  \theta _{3/2}  $. The term in Eq.\ (19)
which is proportional to
$  \theta _{3/2}\theta _{3/2}  $ is zero because the 6j symbol vanishes
(physically
it is related to the vanishing of the $  Q  $ moments in the middle
of a single $  j  $ shell).

\section{Radial densities and momentum distributions}

The radial densities obtained in our potential B
calculations are shown in
various ways in Figs.\ 5-7.
For the sake of simplifiction it will be
assumed in this section that there is a single valence proton
with a binding energy of 0.138 MeV, four tightly bound core protons
(two in the 0p shell with a separation energy of about 6 MeV
and two in the 0s shell with a separation energy of about 16 MeV)
and three tightly bound neutrons (one on the 0p shell with a
separation energy of about 8 MeV and two in the 0s shell with
a separation energy of about 18 MeV). The actual situation is
a little more complicated than this because $  S=1.15  $ and because
there is also some parentage of the protons in $^{8}$B to the first
excited state in $^{7}$Be which will result in some leakage of
the core protons to larger radii. The results here are thus
more qualitative than those given above for $  S_{17}  $.

The normal density $  \rho (r)  $ is shown in Fig.\ 5, the probability
density
$  P(r)=4\pi r^{2}\rho (r)  $ on a log scale in is shown Fig.\ 6, and
the
probability density on a linear scale is shown Fig.\ 7. In all figures
the neutron density is shown by the dashed line, the core proton density
with crosses, and the valence proton density with a
solid line. Note in Figs.\ 6$-$7 that the areas are equal to three, four
and one, respectively.
The valence proton clearly has a large extension, but
whether or not it constitutes a ``halo" or a ``skin" is a question of
semantics.

The valence proton ($  vp  $), core proton ($  cp  $) and neutron
($  n  $) rms radii are:

$$
\sqrt{\langle r^{2}\rangle _{vp}}\, = 4.60 \, {\rm fm} ,
\eqno({21})
$$

$$
\sqrt{\langle r^{2}\rangle _{cp}}\, = 2.39 \, {\rm fm} ,
\eqno({22})
$$

\noindent
  and

$$
\sqrt{\langle r^{2}\rangle _{n}}\, = 2.21 \, {\rm fm} .
\eqno({23})
$$

\noindent
  The total proton rms radius is given by

$$
\sqrt{\langle r^{2}\rangle _{p}}\, = \sqrt{[4\langle r^{2}\rangle
_{cp}+\langle r^{2}\rangle _{vp}]/5}\, = 2.97 \, {\rm fm} .
\eqno({24})
$$

\noindent
  The rms charge radius includes the rms radius of 0.80 fm for the
proton:

$$
\sqrt{\langle r^{2}\rangle _{ch}}\, = \sqrt{[\langle r^{2}\rangle
_{p}+0.64{\rm fm}^{2}]}\, = 3.05 \, {\rm fm} ,       \eqno({25})
$$

\noindent
  and the matter radius is given by:

$$
\sqrt{\langle r^{2}\rangle _{m}}\, = \sqrt{[5\langle r^{2}\rangle
_{p}+3\langle r^{2}\rangle _{n}]/8}\, = 2.71 \, {\rm fm} .
\eqno({26})
$$

Our results can be compared to those of other theoretical calculations.
The results of $\sqrt{ \langle r ^{2}\rangle_{vp} }\,$ = 3.75 fm
obtained by
Riisager and Jensen\,$^{33}$ is much smaller than ours
but they use
an arbitrary potential shape. The reason for their small $  S_{17}  $ is
obvious from Fig.\ 3. The calculations presented in Ref 10 for the
$  Q  $ moment appear to be very close to our results.

The interaction cross sections for $^{8}$B and $^{8}$Li on $^{12}$C have
been calculated with the method of Ref 34. The results with the
finite-range interaction are $\sigma$=843 mb for $^{8}$B and
$\sigma$=820 mb
for $^{8}$Li. These can be compared to the experimental values of
$\sigma$=784(14) mb\,$^{35}$ for $^{8}$B and $\sigma$=768(9)
mb\,$^{36}$ for $^{8}$Li.
The agreement for the magnitudes is as good as can be
expected from the uncertainties in the calculation. However, the
$^{8}$B/$^{8}$Li ratio (in which some of the reaction uncertainties
may cancell) is in excellent agreement between theory and
experiment. The effects of the proton ``halo" in $^{8}$B and the
neutron ``halo" in $^{8}$Li are not large compared with the
classic cases\,$^{34,36}$ of $^{11}$Li, $^{11}$Be and $^{14}$Be.

Recent radiaoactive beam experiments\,$^{37,38}$ have looked at the
momentum distribution of the $^{7}$Be fragments which result from the
break-up of a beam
$^{8}$B on various targets. From these experiments one expects to
measure
the momentum distribution of the most loosely bound protons.
The longitudinal momemtum $  P(k_{z})  $
obtained for the valence proton is shown in
Fig.\ 8. This is obtained by the Fourier transform of the spacial
wave function:

$$
\tilde{\Psi }(\vec{k}) = \frac{1}{(2\pi )^{3/2}} \displaystyle\int
\Psi (\vec{r}) e^{i\vec{k}^{.}\vec{r}} d^{3}r.
       \eqno({27})
$$

\noindent
  where
$  \Psi (\vec{r})=\psi (r) Y_{\ell }(\hat{r})  $,
$  \tilde{\Psi }(\vec{k})=\tilde{\psi }(k) Y_{\ell }(\hat{k})  $, and
where the
radial momentum distribution is given by:

$$
\tilde{y}(k) = \sqrt{\frac{2}{\pi }}\, \frac{i^{-\ell }}{k}
  \displaystyle\int  \psi (r) j_{\ell }(kr) r^{2} \,dr.
\eqno({28})
$$

\noindent
  The logitudinal momentum distribution is given by

$$
P(k_{z}) = \frac{1}{2} \displaystyle\int  \mid \tilde{y}(k) \mid ^{2}
k_{r} dk_{r},       \eqno({29})
$$

\noindent
  where $  k^{2} = k_{r}^{2}+k_{z}^{2}  $.

The calculated momentum distribution has a width of about 150 MeV/c
compared to the experimental value of 81$\pm$6 MeV/c. Given the
good agreement generally found for calculated and obseved neutron
halos\,$^{39}$ this disagreement is puzzling. There is some
discussion in the literature about ways to improve the above
calculation to take into account the peripheral nature of the
reaction.\,$^{39,40}$ In the peripheral direct reaction model one puts
in an
additional cut-off in Eq.\ (29) to exclude the interior part of the
radial distribution which presumably does not contribute because the
cross section coming from that part is dominated by a more violent
reaction where the core ($^{7}$Be in this case) is destroyed. We have
phenomenologically modeled this effect by putting a Fermi shaped
cut-off factor in Eq.\ (28) which has the effect of excluding
the interior out to a radius $  R_{cut}  $ and with a diffuseness
$  a_{cut}  $. We take $  a_{cut}  $=0.65 fm and vary $  R_{cut}  $ to
get about the observed momentum distribtion. This requires
$  R_{cut}  $=5 fm and the results are shown by the dashed line in
Fig.\ 8. The cut-off results in a reduction of $  P(k_{z})  $ at
small momenta by a factor of 2.5 and we have renormalized the
cut-off distribution by this factor in order to show the change in
width. It is already known that the cut-off factor does
not have much effect on the neutron halo momentum
distributions,\,$^{39}$
 and
we have demonstated that even a value as large as
$  R_{cut}  $ = 5 fm
has little effect on the width of the $^{11}$Be neutron halo momentum
distribution. We do not know why we should need $  R_{cut}  $=5 fm,
but we note that this corresponds to the point in Fig.\ 5$-$7 where
the valence proton density falls below the core density.
More work needs to be done to understand these results.

Schwab et al.\,$^{37}$ present an RPA calculation
which goes beyond the
0p shell and which agrees with the shape of the observed momentum
distribution. However, the shape of the wave function beyond about 6 fm
as shown in Fig.\ 4 in their paper appears unrealistic to us. The shape
beyond about 6 fm is entirely determined by the Coulomb and centrifugal
barriers, and their shape is differs from this expectation. Our own
calculation can of course be criticized for staying within the 0p shell.
However, a very recent ``no-core" calculation\,$^{41}$ along the lines
of
those given in Ref 42 which takes into account the lowest six major
shells (21 shell-model orbitals) and up to 4$\hbar\omega$ in excitation
gives
spectroscopic factors which are close to the present 0p shell results.

\section{Conclusions}

Our calculated value of
$  S_{17}(20\, {\rm keV})=26.5\pm 2.0  $ eV-barns
is agreement with the higher values of 25-27 eV-barns
inferred from the ($  p  $,$\gamma$) data of Parker\,$^{4}$ and Kavanagh
et al.,\,$^{5}$ and is higher than the value obtained from the weighted
average
of all experimental data, $  S_{{}17}=22.2\pm 2.3  $ eV-barns,\,$^{43}$
which is the value currently adopted in most solar models.\,$^{44}$
As far as other theoretical predictions are concerned, our current
result for $  S_{{}17}  $ is roughly near the highest of these ($  \sim
 $30
eV-barns)\,$^{45}$ and and much larger than the smallest ($  \sim  $17
eV-barns).\,$^{24,46}$

We note, that in contrast to the common belief (e.g. Ref 46), a small
value of $  S_{17}  $ does not make the solar neutrino problem less
severe. If one takes standard nuclear and solar physics, and standard
neutrino properties, then the best fit\,$^{47}$ of the neutrino fluxes
indicates a suppression in both the $^{7}$Be ($  \phi_{7}  $) and the
$^{8}$B
($  \phi_{8}  $) neutrino fluxes, but the suppression is {\it much}
stronger
in $  \phi_{7}  $. However, a smaller $  S_{17}  $ value {\it alone}
would
make the predicted $  \phi_{7}/\phi_{8}  $ ratio larger, and hence would
exagerage the solar neutrino problem.

\centerline{\bf Acknowledgements}

We would like to acknowledge support from NSF grants 94-03666
and 92-53505. Also we thank Sam Austin and Gregers Hansen for their
helpful comments.

\newpage
\centerline{Figure Captions}

\noindent
  Fig.\ 1: The valence proton rms radius as a function of $\Delta_{o}$.
The three
lines join the points obtained for different value of the diffuseness
$  a  $ for $  R  $=2.0 fm (filled circles),
$  R  $=2.4 fm (open circles) and $  R  $=2.8 fm (squares). The results
for the specific potentials in Table II are shown by the labeled
crosses.

\vspace{ 12pt}
\noindent
  Fig.\ 2: The valence proton density at r=10 fm as a function of
$\Delta_{o}$
(see caption to Fig.\ 1).

\vspace{ 12pt}
\noindent
  Fig.\ 3: The valence proton density at r=10 fm as a function of the
valence proton rms radius (see caption to Fig.\ 1).

\vspace{ 12pt}
\noindent
  Fig.\ 4: The scattering wave function for $^{7}$Li(n,$\gamma$)
obtained with
the standard potential (BI) of Barker\,$^{28}$ (solid line)
together with
the results obtained with the cluster model\,$^{29}$ (dashed line).

\vspace{ 12pt}
\noindent
  Fig.\ 5: The radial density profile for $^{8}$B for the neutrons
(dashed line), the core protons (crosses)
and the valence proton (solid line).

\vspace{ 12pt}
\noindent
  Fig.\ 6: The radial probability distribution for $^{8}$B on a
log scale (see caption to Fig.\ 5).

\vspace{ 12pt}
\noindent
  Fig.\ 7: The radial probability distribution for $^{8}$B on a linear
scale (see caption to Fig.\ 5).

\vspace{ 12pt}
\noindent
  Fig.\ 8: The momentum distribution for the $^{8}$B valence proton. The
solid line corresponds to the full radial wave function and the
dashed line corresponds to the radial wave function cut-off at r=5 fm.

\newpage
\begin{table}
\caption{Values of $\Delta_{o}$, the rms proton radius and $  \rho
(10\, {\rm fm})  $ as a
function of $  R  $ and $  a  $.}
\begin{tabular}{crrrrr}
spin-orbit & $  R  $ (fm) & $  a  $ (fm) & $\Delta_{o}$ (MeV) &
rms (fm) & $  \rho (10\, {\rm fm})  ^{.}$10$^{6}$ (fm$^{-3}$) \\
\tableline
yes & 2.0 & 0.4 & 2.108 & 3.70 & 4.25 \\
yes & 2.0 & 0.6 & 1.947 & 4.14 & 5.81 \\
yes & 2.0 & 0.8 & 1.786 & 4.62 & 8.00 \\
yes & 2.4 & 0.4 & 1.969 & 4.02 & 5.27 \\
yes & 2.4 & 0.6 & 1.840 & 4.40 & 6.85 \\
yes & 2.4 & 0.8 & 1.705 & 4.84 & 9.07 \\
yes & 2.8 & 0.4 & 1.836 & 4.35 & 6.48 \\
yes & 2.8 & 0.6 & 1.735 & 4.68 & 8.10 \\
yes & 2.8 & 0.8 & 1.623 & 5.08 & 10.40 \\
no  & 2.4 & 0.4 & 1.991 & 3.97 & 5.09 \\
no  & 2.4 & 0.6 & 1.851 & 4.38 & 6.74 \\
no  & 2.4 & 0.8 & 1.708 & 4.84 & 9.05 \\
\end{tabular}
\end{table}

\newpage
\begin{table}
\caption{Woods-Saxon potential parameters from the
present analysis (sets A, B and C) compared to those used by
Tombrello\,$^{48}$ and Barker I.\,$^{28}$
Barker II and III correspond to those values Barker needed to
reproduce the $^{7}$Li(n,$\gamma$) cross section.}
\begin{tabular}{crr}
Set & $  R  $ (fm) & $  a  $ (fm) \\
\tableline
A & 2.8 & 0.51 \\
B & 2.4 & 0.69 \\
C & 2.0 & 0.81 \\
Tombrello (T)  & 2.95 & 0.52 \\
Barker BI & 2.39 & 0.65 \\
Barker BII & 0.53 & 0.65 \\
Barker BIII & 2.39 & 0.27 \\
\end{tabular}
\end{table}

\newpage
\begin{table}
\caption{Spectroscopic amplitudes $  \theta _{j}(J_{i},J_{f}=3/2)  $
for $^{8}$B to
$^{7}$Be
from the CKI,\,$^{49}$ Kumar\,$^{50}$ and PTBME\,$^{51}$ interactions.
The order of the states for a given $  J_{f}^{\pi }  $ is indicated by
$  n_{f}  $.}
\begin{tabular}{ccrrr}
$  J_{f}^{\pi },n_{f}  $ & $  j  $ & CKI & Kumar & PTBME \\
\tableline
2$^{ + }$,1 & 3/2 &  0.988 & 0.966 & 0.986 \\
    & 1/2 &  $-$0.237 & $-$0.259 & $-$0.253 \\
1$^{ + }$,1 & 3/2 & 0.567 & 0.606 & 0.552 \\
    & 1/2 & $-$0.352 & $-$0.244 & $-$0.342 \\
3$^{ + }$,1 & 3/2 & 0.581 & 0.555 & 0.565 \\
1$^{ + }$,2 & 3/2 & 0.617 & 0.574 & 0.525 \\
    & 1/2 & 0.840 & 0.861 & 0.859 \\
\end{tabular}
\end{table}

\newpage
\begin{table}
\caption{Spectroscopic factors $  S=S_{1/2}+S_{3/2}  $ for $^{8}$Li and
$^{8}$B.}
\begin{tabular}{ccccccc}
$  J_{f}^{\pi }  $ & $^{8}$B$_{{\rm exp}}$(decay) & $^{8}$Li$_{{\rm
exp}}$(decay) &
$^{8}$Li$_{{\rm exp}}$(reaction) &
CKI & Kumar & PTBME \\
\tableline
2$^{ + }$   &         &         & 0.87(13)        & 1.17 & 1.14 & 1.19
\\
1$^{ + }$   & 0.49(8) &         & 0.48(7) & 0.51 & 0.49 & 0.47 \\
3$^{ + }$   & 0.32(4) & 0.38(7) &      & 0.39 & 0.35 & 0.36 \\
\end{tabular}
\end{table}

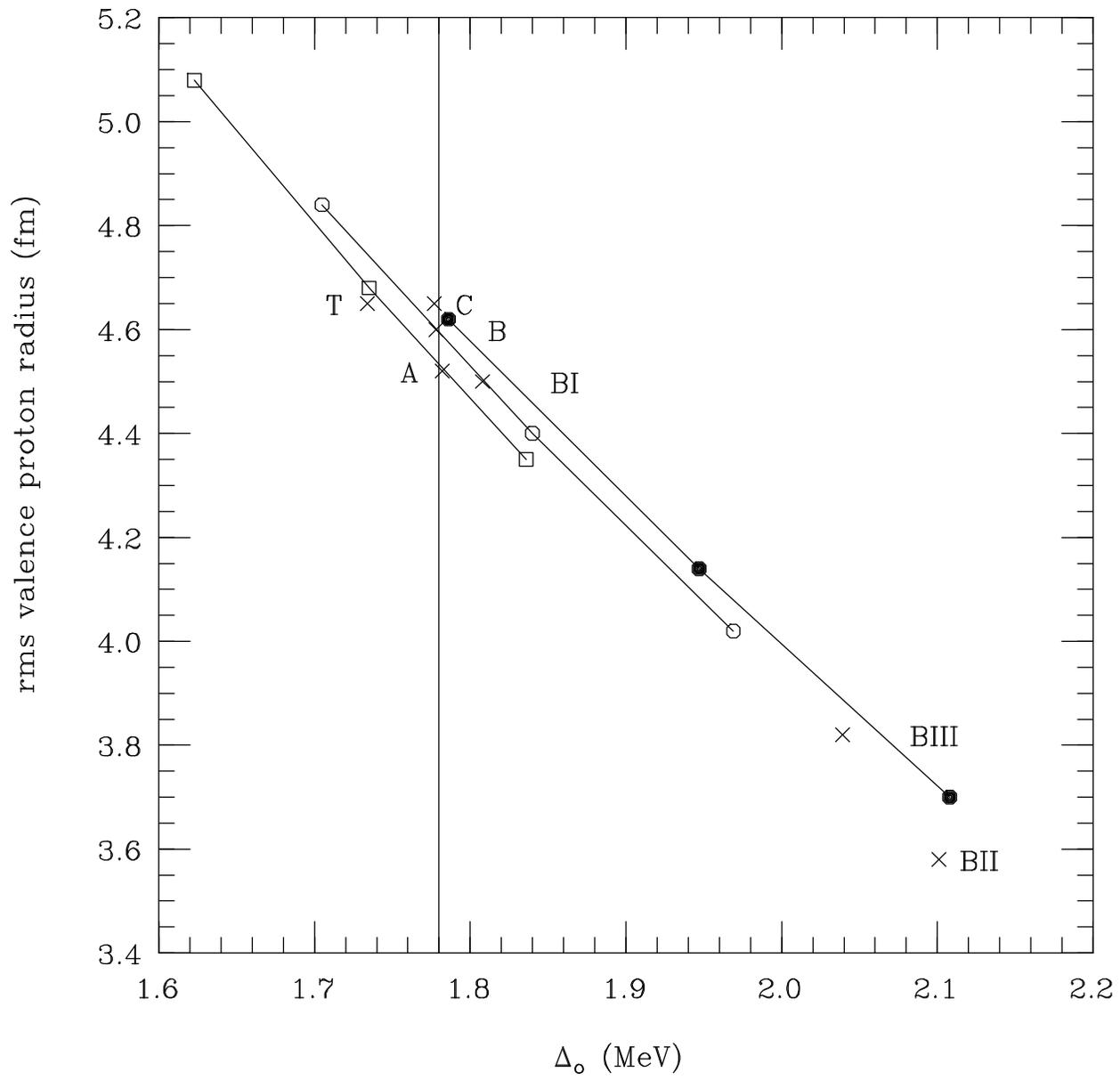
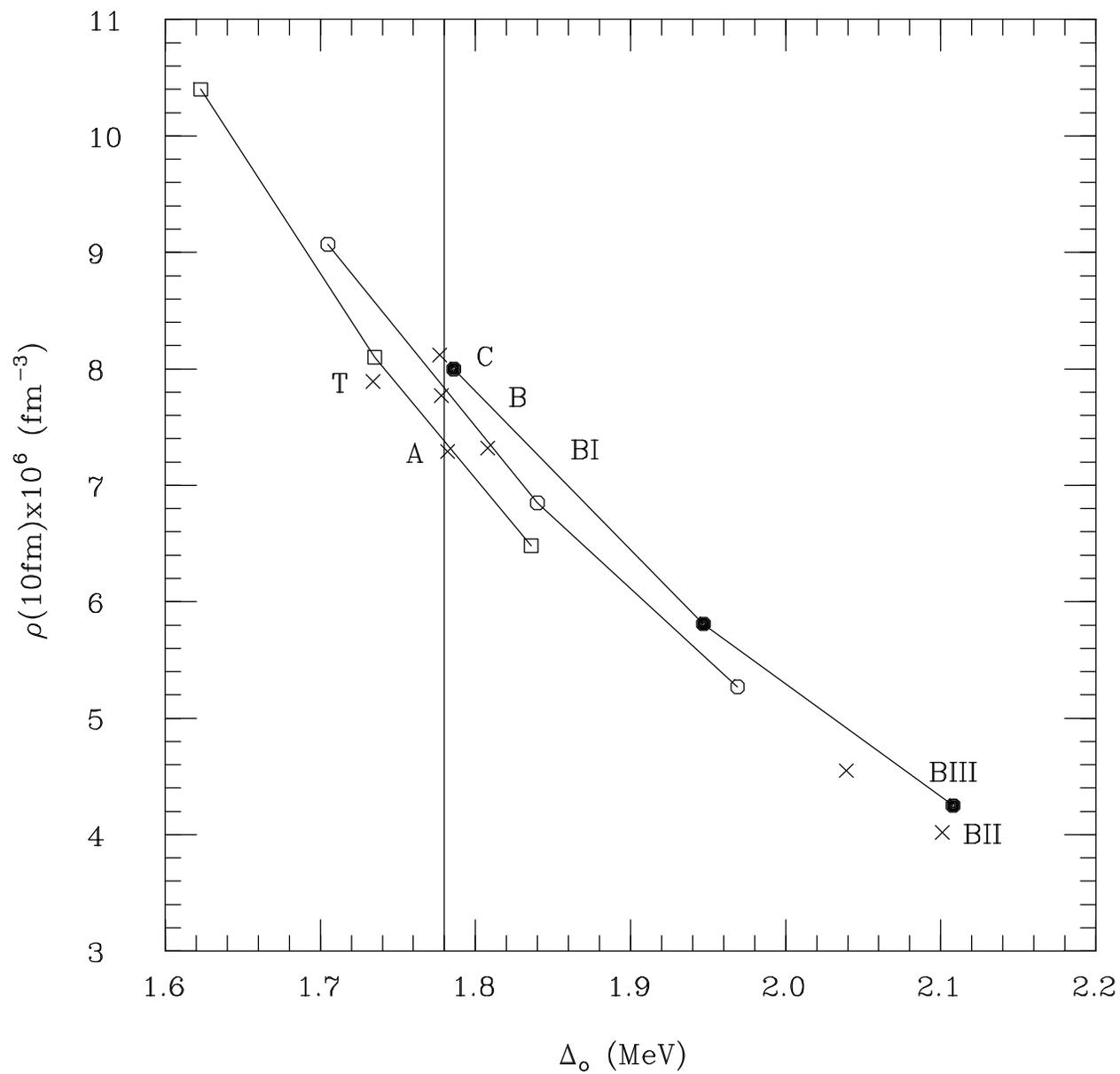
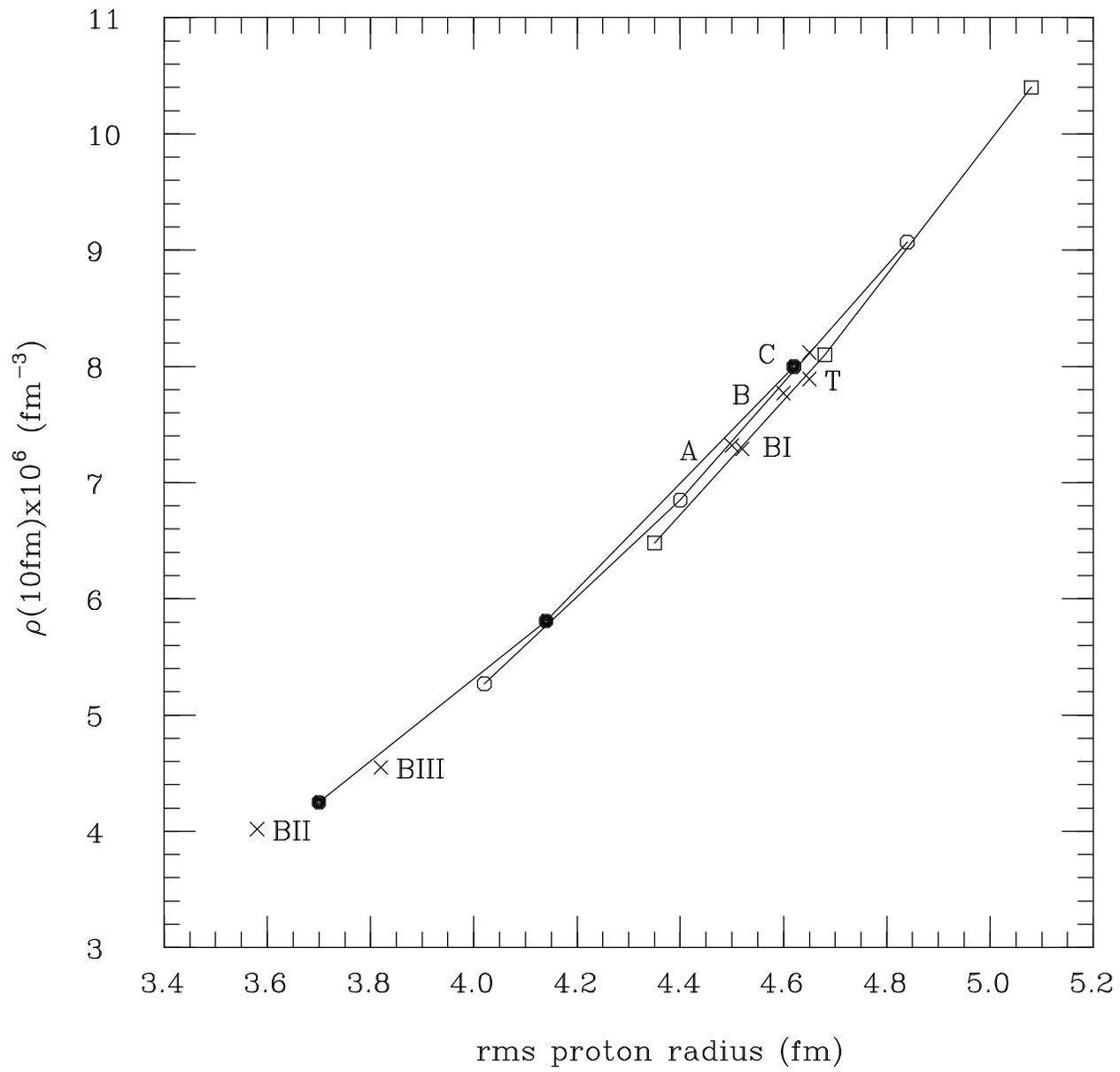
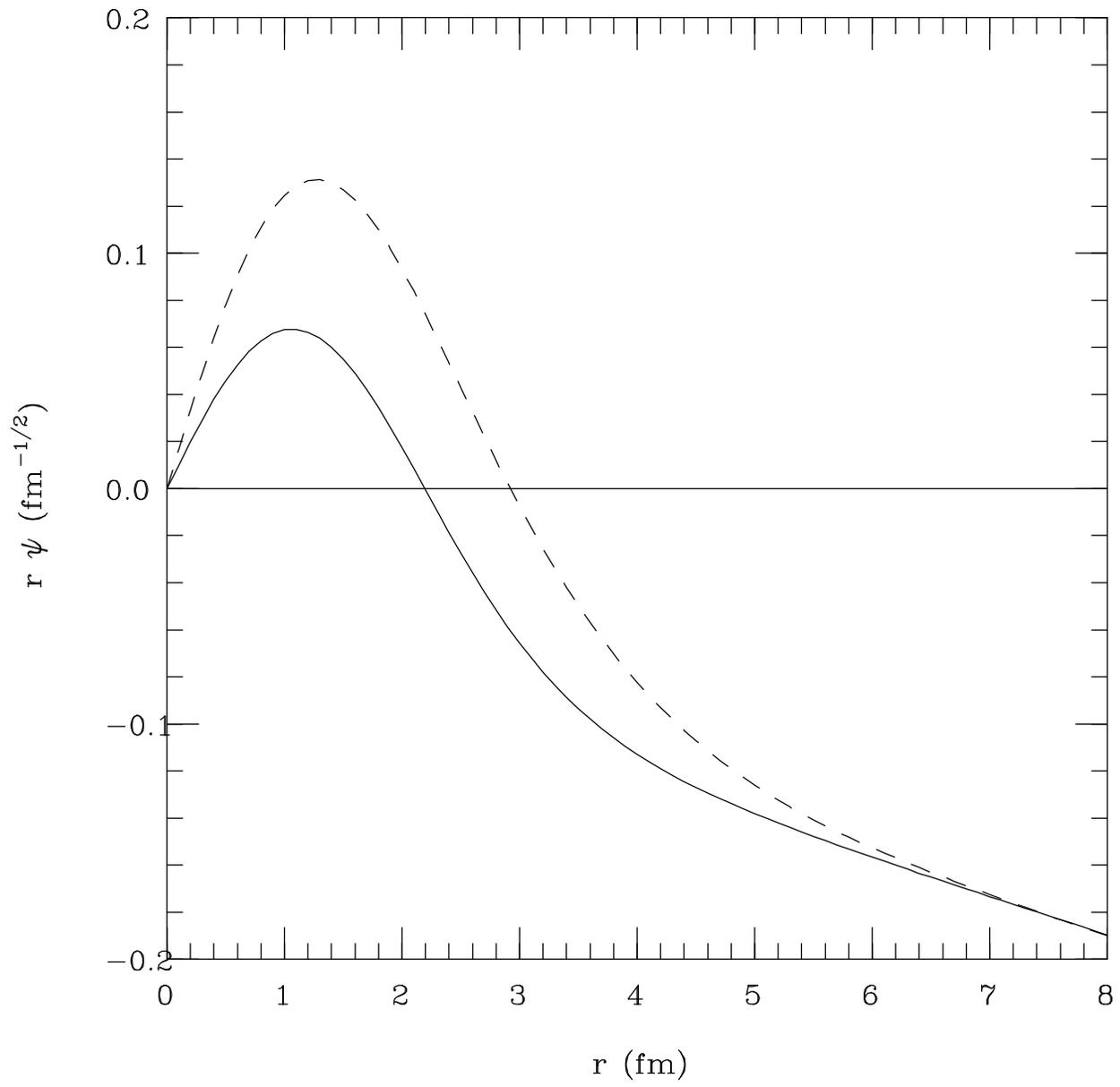
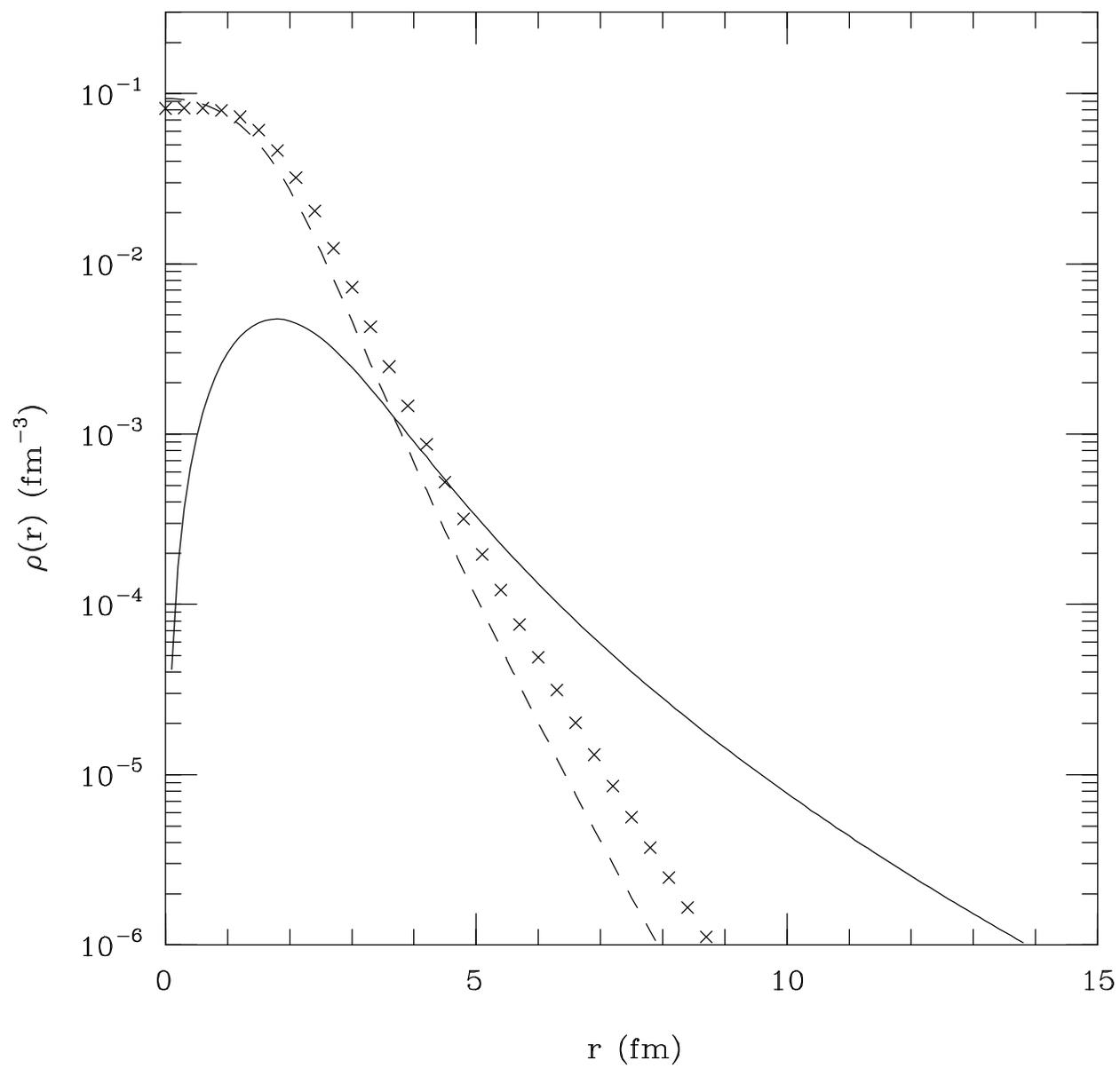
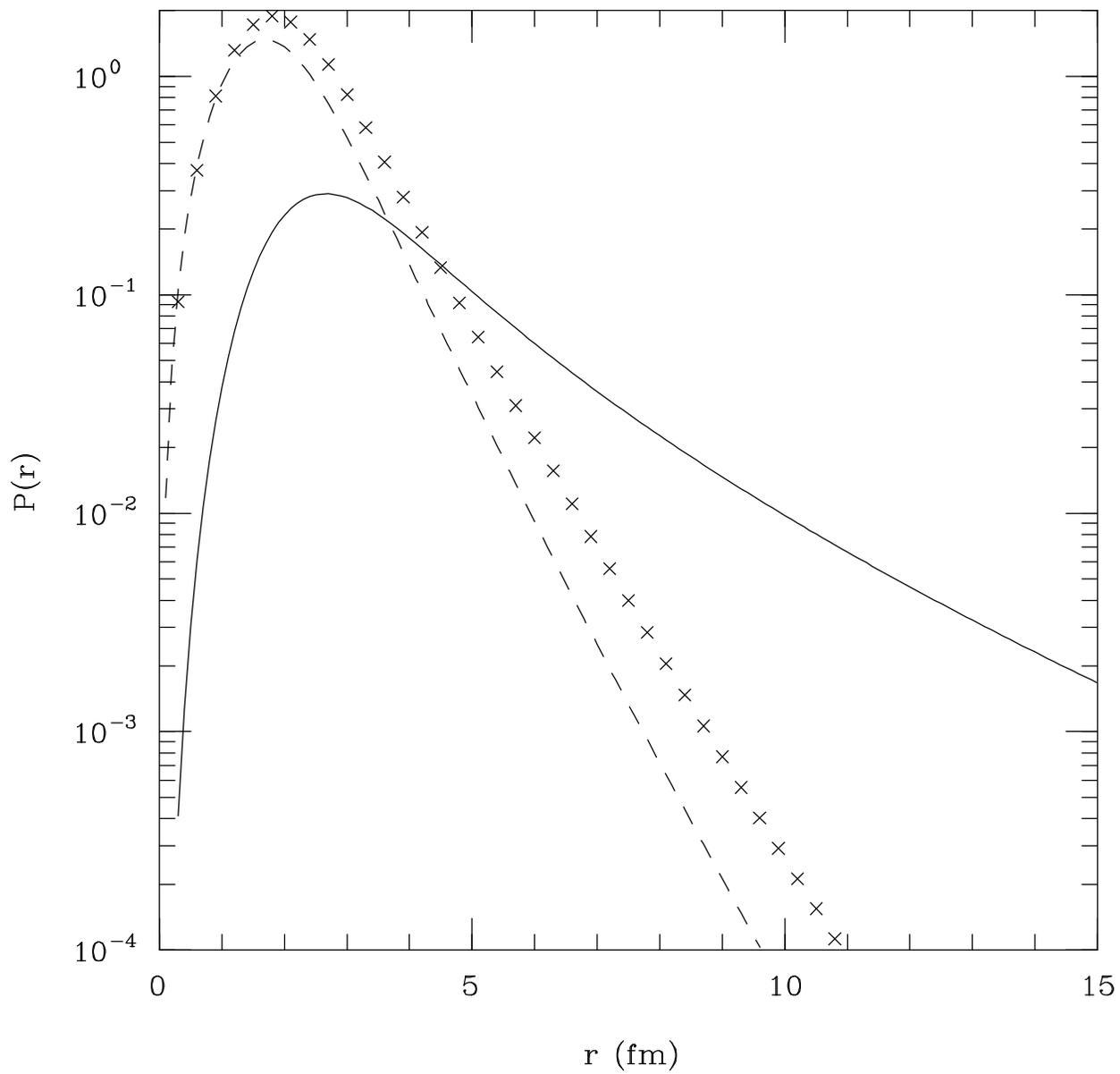
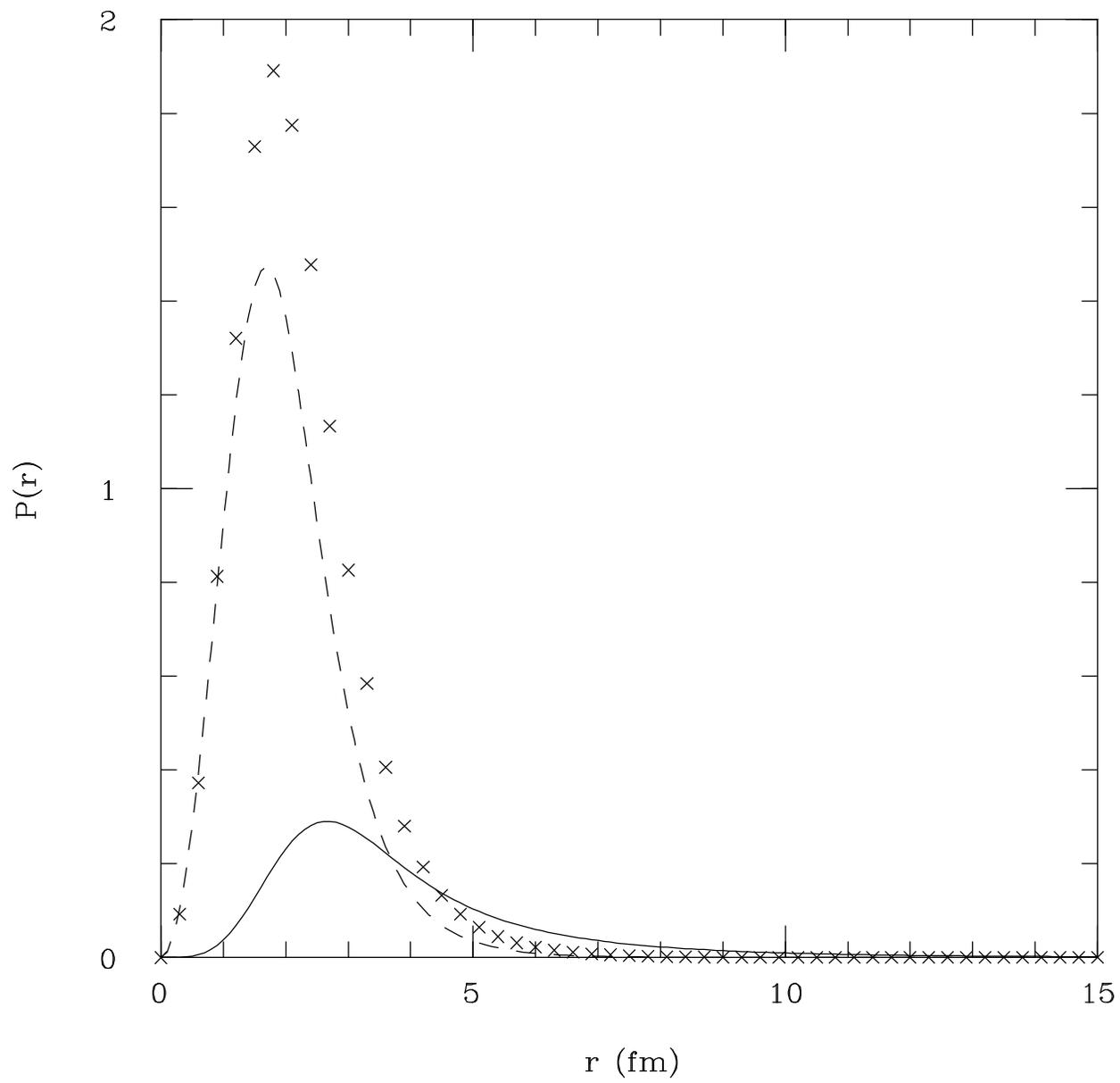
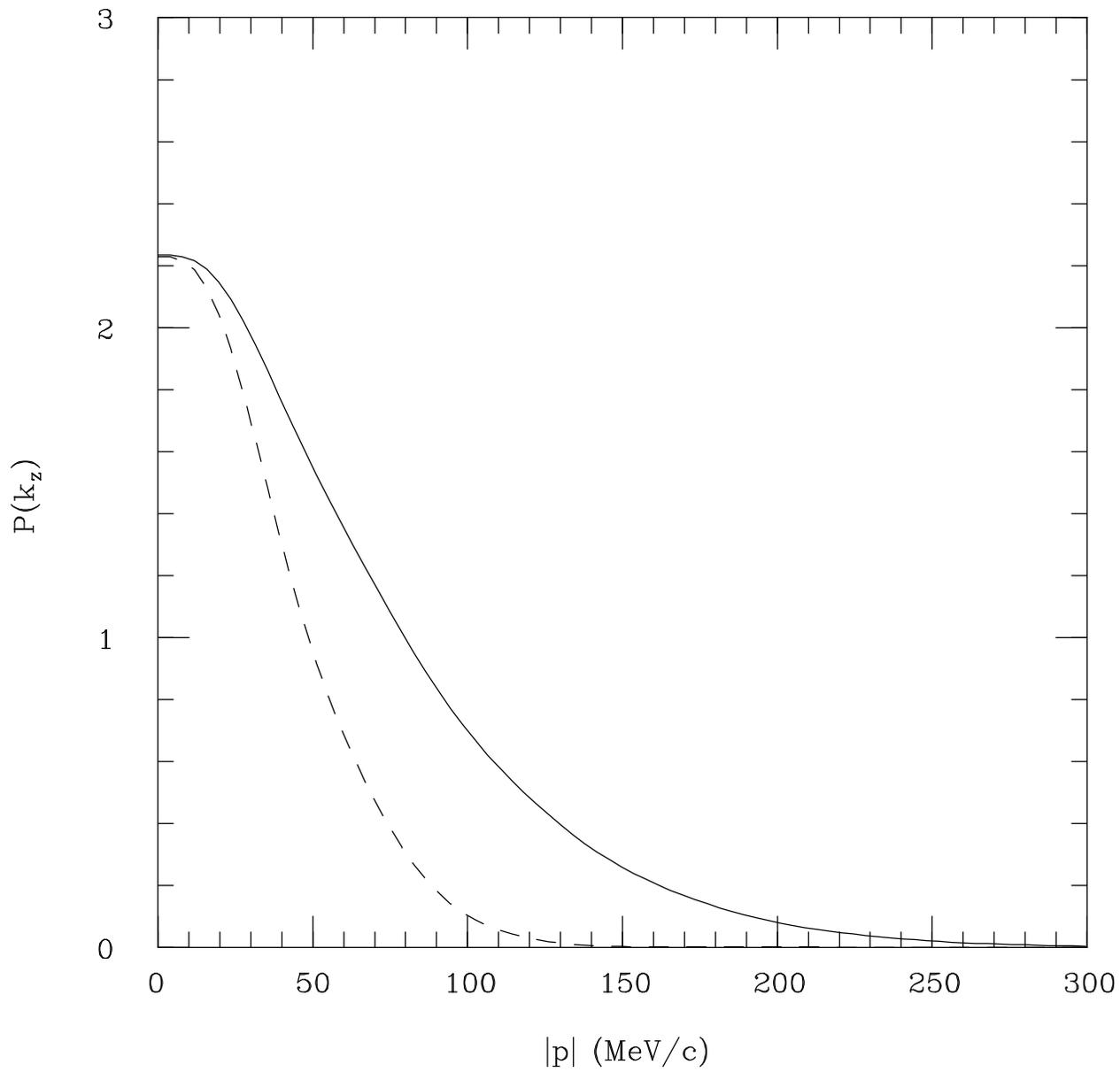
\end{document}